# Rotating Optical Tubes: An Archimedes' Screw for Atoms

Anwar Al Rsheed,[1,*] Andreas Lyras,[1] Omar M. Aldossary[1,2] and Vassilis E. Lembessis[1]

[1]*Department of Physics and Astronomy, College of Science, King Saud University, P.O. Box 2455, Riyadh 11451, Saudi Arabia*

[2]*The National Center for Applied Physics, KACST, P.O. Box 6086, Riyadh 11442, Saudi Arabia*

**ABSTRACT:** The classical dynamics of a cold atom trapped inside a vertical rotating helical optical tube (HOT) is investigated by taking also into account the gravitational field. The resulting equations of motion are solved numerically. The rotation induces a vertical motion for an atom initially at rest. The motion is a result of the action of two inertial forces, namely the centrifugal force and the Coriolis force. Both inertial forces force the atom to rotate in a direction opposite to that of the angular velocity of the HOT. The frequency and the turning points of the atom's global oscillation can be controlled by the value and the direction of the angular velocity of the HOT. However, at large values of the angular velocity of the HOT the atom can escape from the global oscillation and be transported along the axis of the HOT. In this case, the rotating HOT operates as an Optical Archimedes' Screw (OAS) for atoms.

## I. INTRODUCTION

The optical lattice is a standing wave that can be realized by the interference of two counter-propagating beams, which form a periodic pattern of intensity (alternating low and high intensity regions) [1, 2]. However, if the counter-propagating beams have different optical frequencies, then the periodic intensity pattern can be transported, i.e. it can act as an optical atomic conveyor if an atom is trapped at one of the minima of the optical lattice [3].

The trapped atom oscillates around a minimum in the optical conveyor, while the minimum itself is transported. In this case, the optical conveyor acts as a moving frame for the trapped atom, where the speed of the moving frame depends on the difference between the optical frequencies of the beams that form the optical lattice. For example, a one-dimensional optical lattice is formed by two counter-propagating beams and then the atom trapped in this lattice can be transported linearly [3, 4]. The optical Ferris wheel is an optical lattice with cylindrical symmetry. It is formed by two co-propagating Laguerre-Gaussian (L.-G.) beams with opposite helicities $\pm l$. An atom trapped in this lattice can be transported along a circular orbit [5] if the two beams have slightly different frequencies.

Trapped neutral atoms in an optical lattice are one of the most promising candidates for realizing quantum information systems [6]. An optical lattice system was used to trap (or address) single ultra-cold neutral atoms in each potential well to form qubits, which are needed for a quantum register [7]. The goal of a quantum register is to transport selected qubits coherently [8] (the atomic coherence persists while moving the atom back and forth over a macroscopic distance by shifting the optical conveyor [9]) into the interaction zone [10] to create a quantum gate.

The interference of two counter-propagating L.-G. beams with opposite helicities $\pm 1$ produces double helical optical tubes (HOTs) [11, 12]. A static HOT can be produced if the angular frequencies of the L.-G. beams are equal. This scheme has been proposed as an atomic guide along a helical path where the atom oscillates globally between two turning points [13]. However, a slight difference in the angular frequencies of the L.-G. beams produces a rotating HOT [12], which has been used to study the flow of a cold bosonic ensemble (superfluid) that is trapped in the helical pattern [14]. The rotation of the reference frame (the helical pattern) can be used as laboratory equipment with which we can demonstrate the difference between quantum and classical fluids [15]. The rotating helical pattern can also be used as a detector of the slow rotation of an interferometer [16].

The structure of this paper is as follows: The derivation of the dipole potential of an atom inside a rotating HOT is described in Sec. II, while in Sec. III we describe the equations of motion of an atom trapped in the rotating HOT which are integrated numerically. In Sec. IV, we describe the parameters of our numerical calculations and the trajectory of an atom inside a rotating HOT. Moreover, we investigate how the turning points of the trapped atomic motion depend on the angular velocity of the HOT. In Sec. V, we show that under certain conditions a rotating HOT can be used as an Optical Archimedes' Screw (OAS). We were inspired by the famous set up, known since thousands of years ago, with which people were pumping water from rivers and lakes to higher places by just rotating a spiral device. In the same way, by rotating a HOT we can elevate atoms against gravity from a sample of atoms at a lower position. Moreover, in the same section we investigate how radial and global oscillations of the atom depend on the angular velocity of the HOT. Finally, in Sec. VI we present our conclusions.

**II. THE DIPOLE POTENTIAL FOR A TWO-LEVEL ATOM INSIDE A ROTATING HELICAL OPTICAL TUBE**

As we have previously mentioned, the helical optical potential tube is formed by the interference of two counter-propagating L.-G. beams along the $z$-direction, with opposite helicities $\pm l$ and of the same polarization, power and beam waist. If the wave vectors of the counter-propagating beams are almost equal $k_1 \approx k_2 = k$, then the difference between the angular frequencies $\Delta\omega = \omega_1 - \omega_2 \neq 0$ is such that $\Delta\omega = \omega_1 - \omega_2 \ll \omega_1 \approx \omega_2 \approx \omega$. The resultant HOT can be rotated counter-clockwise if $\Delta\omega > 0$ or clockwise if $\Delta\omega < 0$. If we consider that the atom moves close to the beam focus then the dipole potential for a two-level atom trapped inside a rotating HOT is given, in cylindrical coordinates, by:

$$U_d(r', \varphi', z') = \frac{\hbar P \Gamma^2}{2\Delta\pi w_0^2 I_s} \left(u_p^l(r,z)\right)^2 \cos^2(l(\varphi - \Omega_R t) + kz), \tag{1}$$

where $P$ is the power of one of the two beams, $\Gamma$ the excited state spontaneous emission rate, $\Delta$ is the frequency detuning, and $I_s$ the saturation intensity which depends on the specific atomic transition chosen [1]. Finally, $\Omega_R (= \Delta\omega/2l)$ is the angular velocity of the HOT rotation (see Fig. 1) that depends on the sign of $\Delta\omega$.

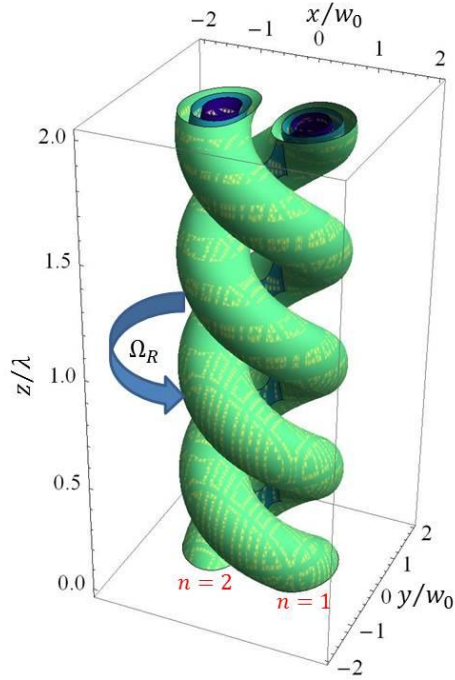

FIG.1. A rotating HOT that is formed by LG beams with $l = 1$ and $p = 0$.

In Eq. (1) the quantity $u_p^l(r,z)$ is given by:

$$u_p^l(r,z) = \sqrt{\frac{p!}{(|l|+p)!}} \frac{w_0}{w(z)} \left(\frac{\sqrt{2}r}{w(z)}\right)^{|l|} exp\left(\frac{-r^2}{w^2(z)}\right) L_p^{|l|}\left(\frac{2r^2}{w^2(z)}\right), \qquad (2)$$

where $L_p^{|l|}(2r^2/w^2(z))$ is an associated Laguerre polynomial and $w(z)\left(= w_0\sqrt{z^2/z_R^2 + 1}\right)$ is the beam waist at position $z$, where $w_0$ is the beam waist at $z = 0$ and $z_R$ is the Rayleigh range, which is equal to $z_R = \pi w_0^2/\lambda$, with $\lambda$ the wavelength of the beam.

## III. THE EQUATIONS OF MOTION OF A SINGLE TWO-LEVEL ATOM INSIDE A ROTATING HOT

The Lagrangian of a cold atom trapped inside the HOT in the laboratory frame, using cylindrical coordinates is:

$$L = \frac{m}{2}\{\dot{r}^2 + r^2\dot{\varphi}^2 + \dot{z}^2\} - U(r,\varphi,z,t), \qquad (3)$$

where $U$ is the total potential energy which is equal to the sum of the dipole potential $U_{dip}$ and gravitational potential, $U_g = mgz$, energies.

We consider that the HOT is rotating about the $z$-axis with angular velocity $\mathbf{\Omega}_R = \Omega_R \hat{\mathbf{z}}$. If we consider that the coordinates of the lab frame are $(r, \varphi, z)$ and the coordinates of a frame attached to the HOT, which is known as the rotating frame, are $(r' = r, \varphi' = \varphi - \Omega_R t, z' = z)$, then the Langrangian in the coordinates of the rotating frame is expressed as follows:

$$L = \frac{m}{2}\{\dot{r}'^2 + r'^2(\dot{\varphi}' + \Omega_R)^2 + \dot{z}'^2\} - U(r', \varphi', z'), \tag{4}$$

where $U(r', \varphi', z')$ is the total potential given as follows:

$$U(r', \varphi', z') = \frac{\hbar P \Gamma^2}{2\Delta \pi w_0^2 I_s}\left(u_p^{|l|}(r', z')\right)^2 \cos^2(l\varphi' + kz') + mgz', \tag{5}$$

where the first term is the dipole potential of a static HOT while the second term is the gravitational potential.

Now we can use the Euler-Lagrange equations to obtain the following equations of motion:

$$m\ddot{r}' = -\frac{\partial U}{\partial r'} + mr'\dot{\varphi}'^2 + 2mr'\Omega_R\dot{\varphi}' + mr'\Omega_R^2 \tag{6a}$$

$$mr'^2\ddot{\varphi}' = -\frac{\partial U}{\partial \varphi'} - 2mr'\dot{r}'\dot{\varphi}' - 2mr'\dot{r}'\Omega_R \tag{6b}$$

$$m\ddot{z}' = -\frac{\partial U}{\partial z'} \tag{6c}$$

The third term in Eq. (6a) and the last term in Eq. (6b) are Coriolis forces, while, the last term in Eq. (6a) is a centrifugal force. These are well known forces that appear because of the rotation of the HOT.

The above three equations of motion are coupled nonlinear differential equations of the second order for which it is very hard to find an exact analytical solution. We can solve these equations numerically using the fourth order Runge-Kutta method.

### IV. THE ATOMIC TRAJECTORY INSIDE THE ROTATING HOT

We consider a $^{85}$Rb atom interacting with the HOT field via the optical transition $5^2S_{1/2} \rightarrow 5^2P_{3/2}$. The interaction is characterized by the following parameters: $\lambda = 780.24 nm$, $I_s = 1.64\, mW/cm^2$, and $\Gamma/2\pi = 5.98 MHz$ [1]. We also consider the following values for the beam power, the detuning and the beam waist: $P = 80mW$ $\Delta = -2.57 \times 10^{13} Hz$, $w_0 = 5\mu m$ [17]. The light field is red detuned with respect to the atomic transition ($\Delta < 0$), so the atom is attracted by the potential well towards the high intensity regions. The initial position of the cold atom can be chosen at the maximum intensity location $\left(\sqrt{l/2}\, w_0, 0, 0\right)$ inside the tube with index $n = 1$, as shown in Fig. 1.

In Figs. 2 (a) and (b) we present the trajectories of a $^{85}$Rb atom, initially at rest, in the rotating HOT frame of reference. On the other hand, in Figs. 3 (a) and (b) we present the trajectories of the $^{85}$Rb atom, initially at rest, in the lab frame of reference. In Figs. 3 (a) and (b) we can see that the tapped atom in the lab frame has faster azimuthal rotation velocity equal to $\dot{\varphi} + \Omega_R$. Moreover, the atomic trajectories in Fig. 2 (a) and Fig. 3 (a) show that the $^{85}$Rb atom is elevated when the HOT rotates counter-clockwise with $\Omega_R = 70KHz$. On other hand, the $^{85}$Rb atom in Fig. 2 (b) and Fig. 3 (b) follows a downward path when the HOT is rotated clockwise with $\Omega_R = -70KHz$. Consequently, the, initially at rest, cold atom will acquire an induced clockwise azimuthal velocity if it is trapped inside a counter-clockwise rotating HOT, while it will acquire an induced counter-clockwise azimuthal velocity if it is trapped inside a clockwise rotating HOT. In other words, the angular momentum that is transferred from the rotating HOT to the atom is directed opposite to the angular velocity of rotation of the HOT, in agreement with the prediction by Bekshaev et al [18].

The induced motion of the atom that is trapped inside a rotating HOT can be explained in terms of the inertial forces. Initially, the atom inside a rotating HOT will experience a centrifugal force $F_r = mr_o\Omega_R^2$. Consequently, the atom will move along the radial direction. After an infinitesimal time interval $\Delta t$, the atom will have a nonzero radial velocity $\dot{r}' = r_o\Omega_R^2\Delta t$. As a result, a Coriolis force $F_\varphi = -2m\dot{r}'\Omega_R$ will be exerted on the atom along the azimuthal direction which drives the atom to rotate in a direction opposite to that of the angular velocity $\Omega_R$ of the HOT. Finally, the atom will have a global motion along the HOT due to the coupling between the motions along the z- and the azimuthal directions [13].

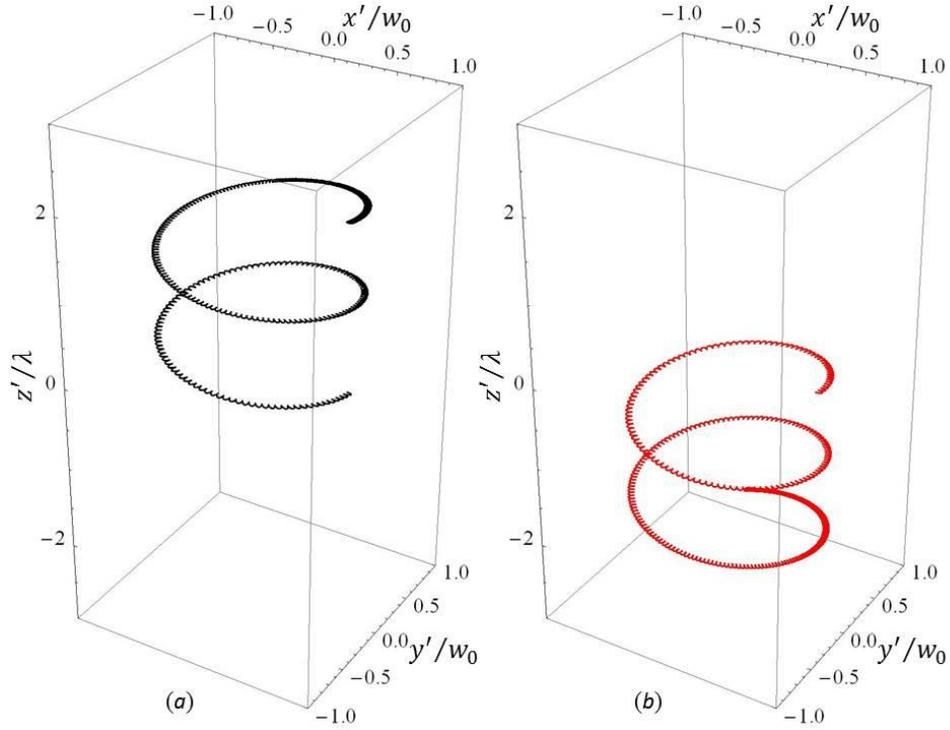

FIG. 2. The trajectory of a $^{85}$Rb atom with respect to the HOT frame of reference: (a) $\Omega_R = 70 kHz$ represented by the black line (b) $\Omega_R = -70 kHz$ represented by the red line.

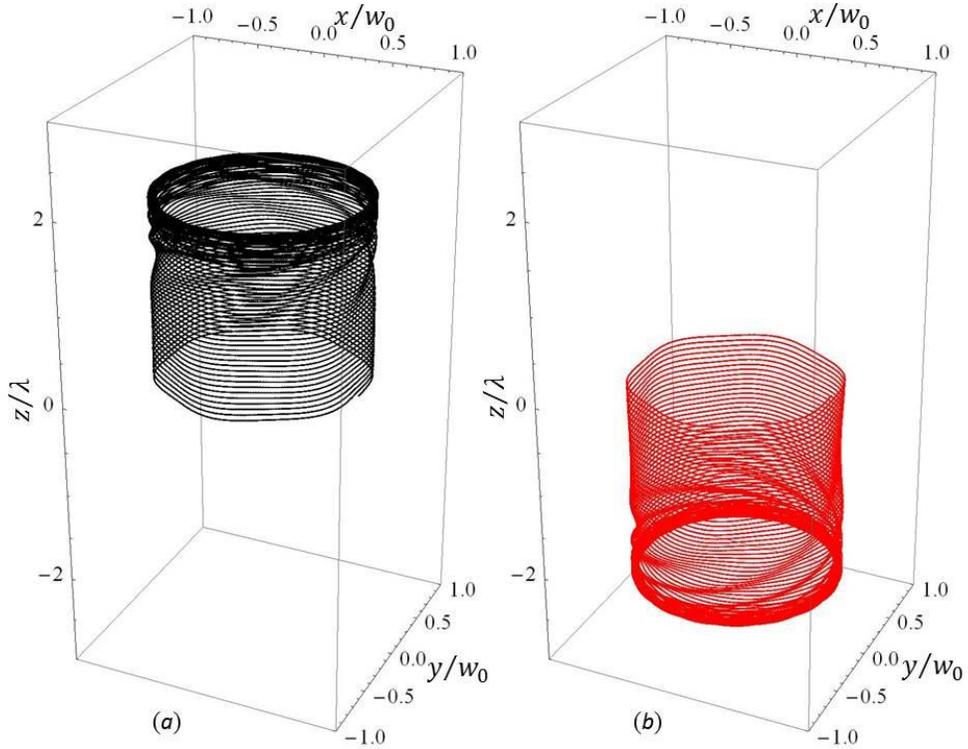

FIG. 3. The trajectory of a $^{85}$Rb atom with respect to the lab frame of reference: (a) $\Omega_R = 70 kHz$ represented by the black line (b) $\Omega_R = -70 kHz$ represented by the red line.

We must point out that it is not possible to have an atom at rest. The atom no matter how cold it will always have a non-zero velocity. Also when the atom is very cold a semi-classical description of its motion is not valid and we must treat its center-of-mass motion quantum mechanically. Therefore, the above discussion is preliminary. In the next section we consider the case of an atom with non-zero initial velocity such that a semi-classical approximation of its motion is justified.

We consider now the case where the atom has an initial velocity such that a semi-classical description of its gross motion can be justified [13]. In Fig. 4 we see the variation of the atomic elevation as a function of time for many different angular rotational velocities of the HOT $\Omega_R$. This figure shows that the turning point in the atomic elevation depends on the value and direction of $\Omega_R$. Specifically, the value of the upper turning point of the trapped atom (with $\dot{\varphi}_o < 0$) increases as the value of $\Omega_R$ increases, if $\Omega_R$ is counter-clockwise, or decreases if $\Omega_R$ is clockwise. This is so because the transferred angular momentum from the rotating HOT is opposite to the direction of the atom's rotation. Moreover, if $\Omega_R \approx 97 krad/s$ in the clockwise direction and $\dot{\varphi}_o = -14.14\ krad/s$, then the atom can be localized within a size $\Delta z \ll \lambda$ around the minima $z = 0$ which is represented by the brown dotted line in Fig. 4.

In Fig. 5 we present the variation of the value of the first turning point in the atomic motion along the z-axis as a function of the angular rotational velocity of the HOT $\Omega_R$, for different values of the initial atomic velocity. From the shape of the different curves we can conclude that the effect of the initial velocity is only to shift the curve up (if $\dot{\varphi}_o$ is clockwise) or down (if $\dot{\varphi}_o$ is counter-clockwise). This figure confirms the statement that the rotating HOT carries angular momentum opposite to the direction of the atom's rotation which is transferred to the atom during the interaction between the HOT and the trapped atom.

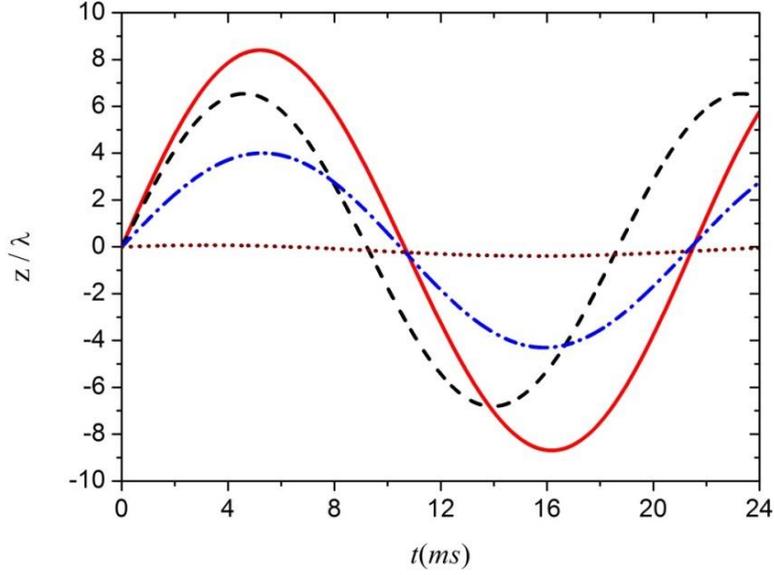

FIG. 4. The variation of a $^{85}$Rb atom elevation (the initial velocity of the atom is $(v_x = 5\,cm/s, v_y = -5\,cm/s, v_x = 0)$ for different angular rotation velocities of the HOT $\Omega_R$: $70\,krad/s$ (red solid line), static HOT (black dashed line), $-70\,krad/s$ (blue dashed-dotted line), $-97\,krad/s$ (brown dotted line).

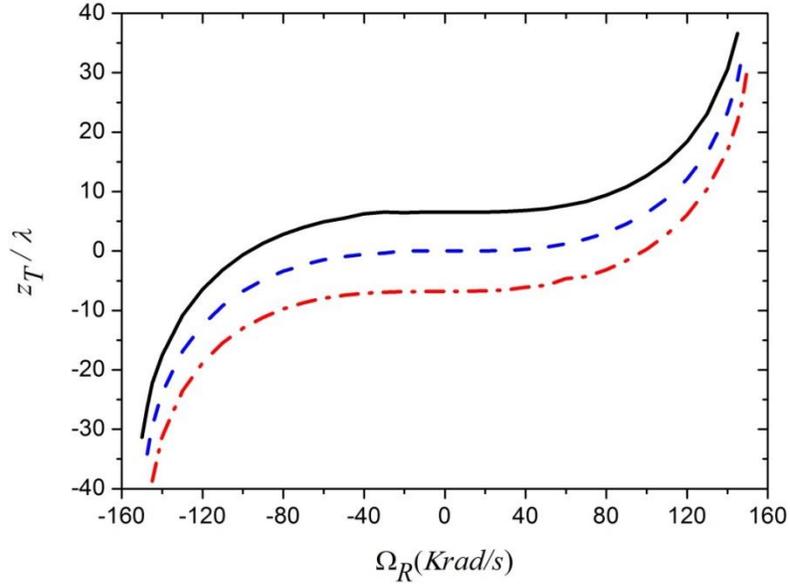

FIG. 5. The variation of the first turning point of a $^{85}$Rb atom as a function of the angular velocity of the HOT for different initial atomic velocities: $(v_x = 5\,cm/s, v_y = -5\,cm/s, v_x = 0)$ black line, $(v_x = -5\,cm/s, v_y = 5\,cm/s, v_x = 0)$ red dashed-dotted line, and $(v_x = 0, v_y = 0, v_x = 0)$ blue dashed line ($w_0 = 5\mu m$ and $l = 1$).

# V. THE ROTATING HOT AS AN OPTICAL ARCHIMEDES' SCREW (OAS) FOR ATOMS

We will now demonstrate that the rotating HOT can be used for elevating atoms if we make a proper choice of the involved parameters. We must note that elevation has two prerequisites: first the atom must be able to "escape" from the oscillations along the vertical z-direction and, second, it must, simultaneously, be kept trapped in the potential tube without escaping along the radial $r$-direction. In Fig. 6 we present the vertical displacement of the atom as a function of interaction time. We see that the atom, for the chosen initial velocity, can be elevated along the z-axis when the HOT rotates at angular velocities greater than 146 $krad/s$ counter-clockwise. It can also move downwards when the HOT rotates at angular velocities greater than 150 $krad/s$ clockwise. For other values of the rotational angular velocity it clearly performs an oscillation along the z-axis, which means that it remains trapped in this direction. From Fig.7 we can check the state of the atomic motion in the radial direction. We see that if the HOT rotates with $\Omega_R$ greater than 279 $krad/s$ clockwise or 307 $krad/s$ counter-clockwise, then the atom will escape from the helical tube along the radial direction.

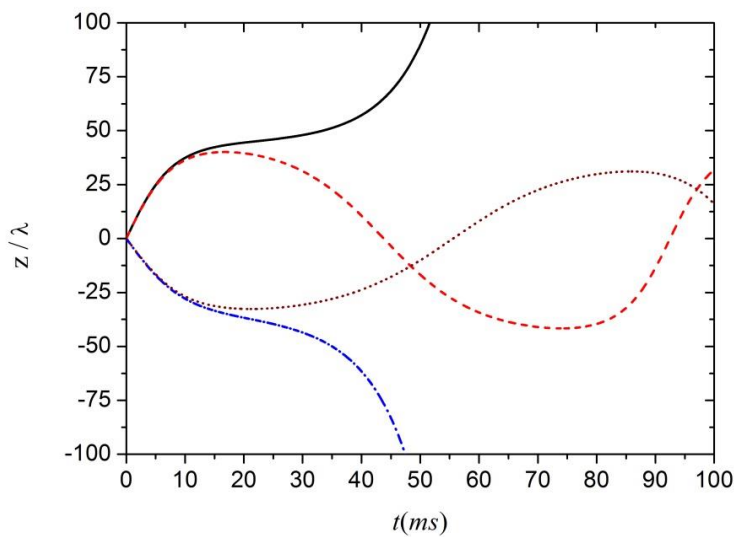

FIG. 6. The variation of a $^{85}$Rb atom elevation with initial velocity $(v_x = 5\ cm/s, v_y = -5\ cm/s, v_x = 0)$ for different angular rotation velocities of the HOT: $147 krad/s$ (black solid line), $146 rad/s$ (red dashed line), $-151 krad/s$ (blue dashed-dotted line), $-150 krad/s$ (brown dotted line) ($w_0 = 5\mu m$ and $l = 1$).

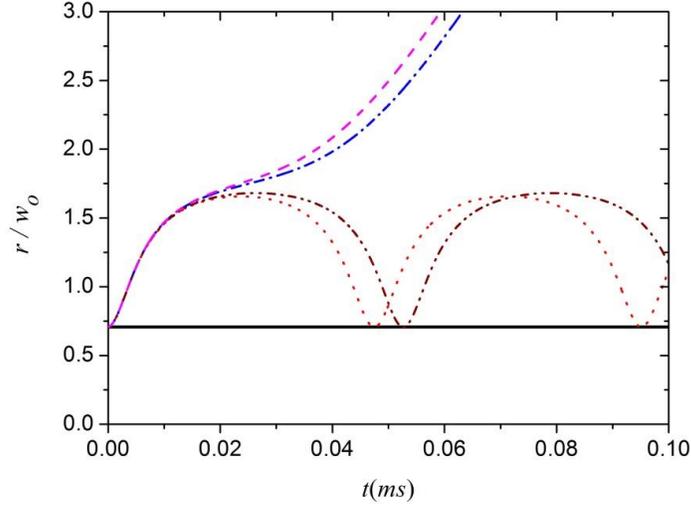

FIG. 7. The variation of a $^{85}$Rb atom radial wiggling with initial velocity $(v_x = 5\ cm/s, v_y = -5\ cm/s, v_x = 0)$ for different angular rotation velocities of the HOT: $307\ krad/s$ (red dotted line), $308\ krad/s$ (blue dash-dotted line), $-279\ krad/s$ (brown dash-dotted-dotted line), $-280\ krad/s$ (dashed pink line) (the black solid line represent to radial minima of the potential; $w_0 = 5\mu m$ and $l = 1$).

Let us now try to understand analytically the effect of the angular rotation of the HOT $\Omega_R$ on the frequencies of the global oscillation. We introduce the Frenet coordinates $\rho, \xi,$ and $\nu$, defined as:

$$\rho = r - \sqrt{\frac{|l|}{2}} w(\xi), \tag{7a}$$

$$k\nu = l\varphi + kz, \tag{7b}$$

$$\xi = -\varphi = \frac{z}{h}, \tag{7c}$$

where $2\pi h (= |l|\lambda)$ is the helical pitch of the HOT, $w(\xi) = w_0\sqrt{\alpha\xi^2 + 1}$ the beam waist as function of the helical parameter $\xi$ and $\alpha (= h^2/z_R^2)$.

The Taylor expansion of the dipole potential around its minima, using the Frenet coordinates, is as follows:

$$U(\rho, \xi) = -\varepsilon\left\{-k^2\nu^2 + (1 - \alpha\xi^2 + \alpha^2\xi^4 + \cdots) - (1 - 2\alpha\xi^2 + 3\alpha^2\xi^4 + \cdots)\frac{4\rho^2}{w_0^2} + \cdots\right\} \tag{8}$$

Note that we keep only one term from the expansion of $\cos^2(k\nu)$ because $k\nu \ll 1$. The constant $\varepsilon$ in Eq. (8) is:

$$\varepsilon = -\frac{\hbar P \Gamma^2}{2\Delta \pi w_0^2 I_s} \frac{|l|^{|l|}}{|l|!} e^{-|l|}, \tag{9}$$

and represents the depth of the dipole potential [13].

The relation between cylindrical unit vectors and Frenet unit vectors (one of them is tangent to the helix, the second normal to the helix, and the third one is binormal to the helix) is [19]:

$$\hat{r} = \hat{\rho}, \tag{10a}$$

$$\hat{z} = \sin(\theta)\,\hat{\xi} + \cos(\theta)\,\hat{v}, \tag{10b}$$

$$\hat{\varphi} = -\cos(\theta)\,\hat{\xi} + \sin(\theta)\,\hat{v}, \tag{10c}$$

where $\theta(=\tan^{-1}(h/r))$ is the pitch angle. In our calculation of the kinetic energy of the atom inside the HOT we should consider the following: $r \gg h$, $w'(\xi) \ll w(\xi)$ and $\xi < \alpha^{-1}$ (in Fig. 5 we can see the maximum value of the turning point that the atom can reach and still be trapped globally between two turning points is $|z_T| = 38\lambda$ and recalling that $z_R = 129\lambda$, then $|z_T|/z_R \approx 0.3$, which means that $z < z_R$). The kinetic energy of the atom can be approximated as follows:

$$T(\rho, \xi, \dot{\rho}, \dot{\xi}, \dot{v}) = \frac{m}{2}\left\{\dot{\rho}^2 + \left(\sqrt{\frac{|l|}{2}}w(\xi) + \rho\right)^2 (\Omega_R - \dot{\xi})^2 + h^2(\dot{\xi}^2 - \Omega_R^2)\right.$$

$$\left. + (h\Omega_R + \dot{v})^2 + 2\sqrt{\frac{|l|}{2}}w'(\xi)\dot{\rho}\dot{\xi}\right\}. \tag{11}$$

The equation of motion along $\xi$ can be formulated from the Euler-Lagrange equation as follows [20] (we present only the terms that include $\Omega_R$ and show its effect on the global oscillations):

$$2\beta(\xi, \rho)\ddot{\xi} + 2\beta\left\{\omega_s^2 - \frac{\lambda^4}{4\pi^4}\frac{|l|^2}{w_0^4}\Omega_R^2\right\}\xi = -mgh + m\sqrt{\frac{|l|}{2}}w'(\xi)\Omega_R^2\,\rho$$

$$+ 2m\left(\sqrt{\frac{|l|}{2}}w(\xi) + \rho\right)\Omega_R\dot{\rho} + non-linear\ terms, \tag{12}$$

where $\beta(\xi, \rho) = m\left\{\left(\sqrt{|l|/2}\,w(\xi) + \rho\right)^2 + h^2\right\}/2$.

This equation of motion for the global oscillation can show easily the effect of the inertial forces on the global motion because the term $m\sqrt{|l|/2}\,w'(\xi)\Omega_R^2\,\rho$ corresponds to the centrifugal force, while the term $2m\left(\sqrt{|l|/2}\,w(\xi)+\rho\right)\Omega_R\dot{\rho}$ corresponds to the Coriolis force. The Coriolis force term reveals the effect of the sign of $\Omega_R$ on the global motion: the atom is elevated when the HOT rotates counter-clockwise while it moves downwards when the HOT rotates clockwise, which confirms the results in Fig. 2.

From Eq. (12), which cannot be solved analytically, we can draw some qualitative conclusions, which support the findings of our numerical calculations. The frequency of the oscillations along the helix is mainly determined by the term in the curly brackets of Eq. (12). So we could say that this term defines a frequency given by:

$$\omega_s' \approx \sqrt{\omega_s^2 - \frac{\lambda^4}{4\pi^4}\frac{|l|^2}{w_0^4}\Omega_R^2}, \tag{13}$$

where $\omega_s = \sqrt{\alpha\varepsilon/\beta}$ is the frequency of the global oscillation of an atom that is trapped in a static HOT [13] and $\beta = m\{|l|w_0^2/2 + h^2\}/2$. This expression can give us an explanation for the decay of the effective global frequency $\omega_s'$ with $\Omega_R$ (as depicted in Fig. 8). Moreover, Eq. (13) shows that the effective global frequency does not depend on the sign of $\Omega_R$, because it depends on $\Omega_R^2$, which explains why the effective global frequency graph has parabolic shape with $\Omega_R$ as is clearly shown in Fig. 8. The value of $\omega_s'$ when $\Omega_R = 0$ as shown in the figure is, for the same parameters, exactly equal to the values given in [20, 13] where we investigated the atom dynamics in a static HOT.

Moreover, in Fig. 8 we can see the variation of the effective radial frequency $\omega_r'$ with $\Omega_R$. It is evident that also the effective radial frequency does not depend on the sign of $\Omega_R$. It was very hard to find an analytic expression for $\omega_r'$ as a function of $\Omega_R$, because the atomic motion along the $r$-direction couples strongly with the motion along the $z$-direction. This coupling can be seen clearly in Fig. 8: the value of the effective radial frequency $\omega_r'$, initially increases slowly with $\Omega_R$, it attains its maximum value when the values of $\omega_s'$ becomes zero and then decreases sharply with $\Omega_R$ following a similar dependence as $\omega_s'$.

The important feature in Fig. 8 is that there are values of the HOT angular frequency $\Omega_R$ for which the oscillation frequencies along the axial and radial direction, $\omega'_s$ and $\omega'_r$ respectively, become equal to zero. These values are denoted with the symbols $\Omega_R^s$ and $\Omega_R^r$, respectively, and they are very crucial in our analysis since they determine the range of HOT rotation frequency values for which we can achieve elevation of atoms along the axial direction.

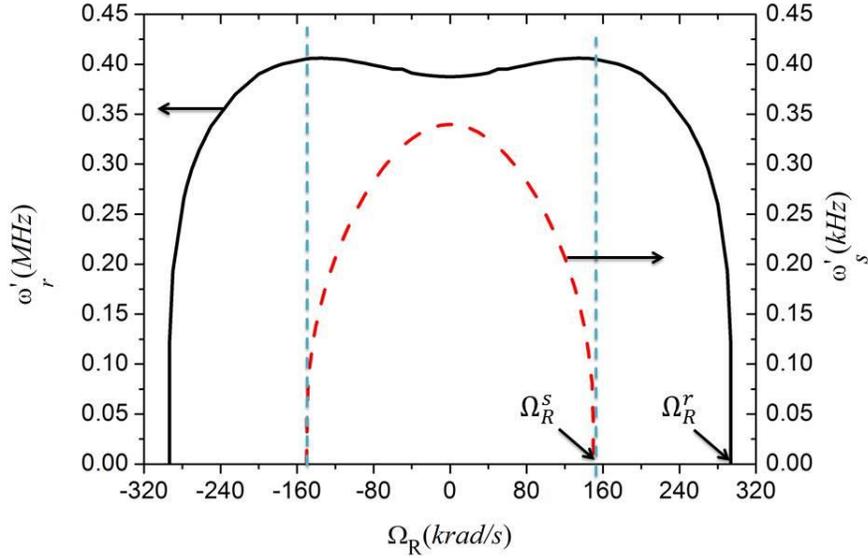

FIG. 8. The variation of $\omega'_s$ (red dashed line) and $\omega'_r$ (black solid line) of a trapped $^{85}$Rb atom as a function of $\Omega_R$. The atom started the motion from rest, the beam waist of LG beams is $w_0 = 5\mu m$, and the orbital angular momentum number is $l = 1$.

From the above figure we see that the atomic radial frequency $\omega'_r$ is always larger than the atomic global frequency $\omega'_s$. However, we can choose the global and radial frequency resonances to use the HOT as an Optical Archimedes' Screw (OAS). If the HOT rotates within the following range of angular velocities:

$$|\Omega_R^s| < |\Omega_R| < |\Omega_R^r|, \qquad (14)$$

then the atom can be elevated to any desired height along z-axis while simultaneously remaining trapped in the radial direction. Moreover, the atom can be transported upwards or downwards along the HOT by changing the direction of rotation of the HOT. The values of $\Omega_R^s$ and $\Omega_R^r$ can be controlled by changing the dipole potential parameters such as the orbital angular momentum number $l$ (see Fig. 9), the beam waist $w_0$ (see Fig. 10), the beam power $P$ (see Fig. 11), and the detuning $\Delta$ (see Fig. 12). All Figs. 9, 10, 11, and 12 show three regions: the region below the red dashed line including the values of $\Omega_R$ for which the atom inside the rotating HOT still undergoes a global oscillation. The region between the black and red dashed lines corresponds to values of $\Omega_R$ for which the rotating HOT can be used as an OAS. Finally, the region above the black dashed line corresponds to values of $\Omega_R$ for which the atom will gain enough kinetic energy to escape from the rotating HOT. Additionally, the figures show that $\Omega_R^s$ and $\Omega_R^r$ can have large values if the dipole potential becomes stronger and smaller ones if the dipole potential becomes weaker.

Moreover, the range between the values of $\Omega_R^s$ and $\Omega_R^r$ can be controlled by changing the dipole potential parameters such as the orbital angular momentum number $l$, the beam waist $w_0$, the beam power $P$, and the detuning $\Delta$. This range becomes wider when the dipole potential becomes stronger and narrower when the dipole potential becomes weaker, as evident in Figs. 9, 10, 11, and 12.

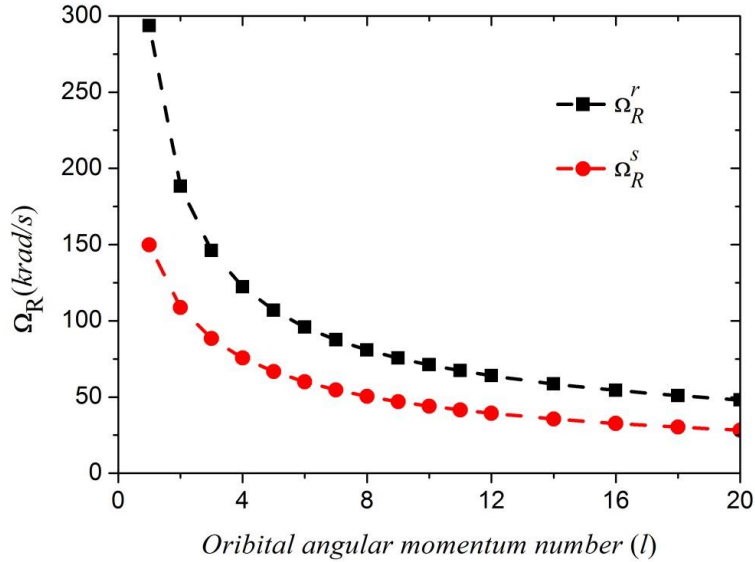

FIG. 9. The variation of $\Omega_R^r$ and $\Omega_R^s$ as a function of the orbital angular momentum number $l$ for a $^{85}$Rb atom initially at rest, where $w_0 = 5\mu m$, $P = 80 mW$, and $\Delta = -25.7 THz$.

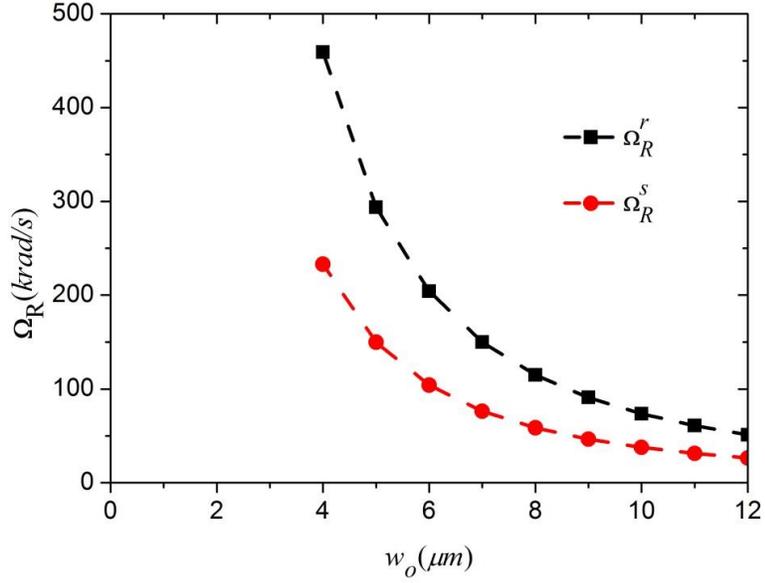

FIG. 10. The variation of $\Omega_R^r$ and $\Omega_R^s$ as a function of the beam waist $w_0$ for a $^{85}$Rb atom initially at rest, where $l = 1$, $P = 80 mW$, and $\Delta = -25.7 THz$.

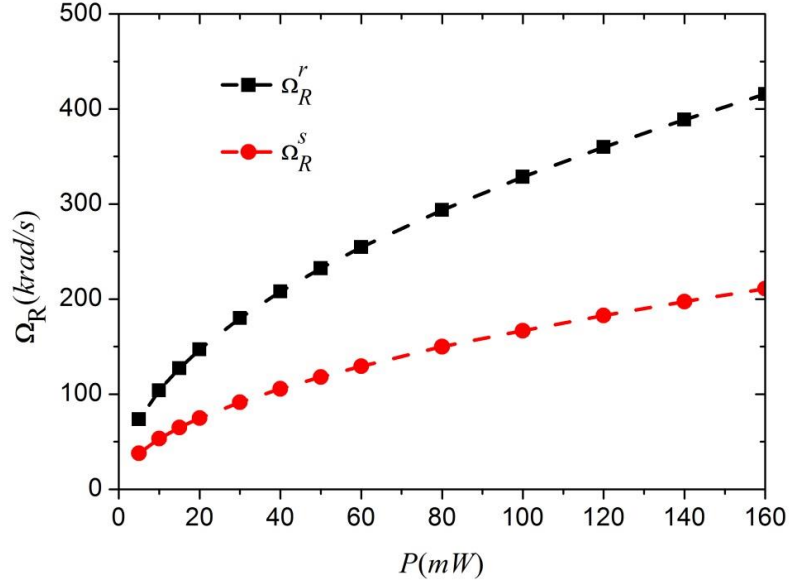

FIG. 11. The variation of $\Omega_R^r$ and $\Omega_R^s$ as a function of the beam power $P$ for a $^{85}$Rb atom initially at rest, where $l = 1$, $w_0 = 5 \mu m$, and $\Delta = -25.7 THz$.

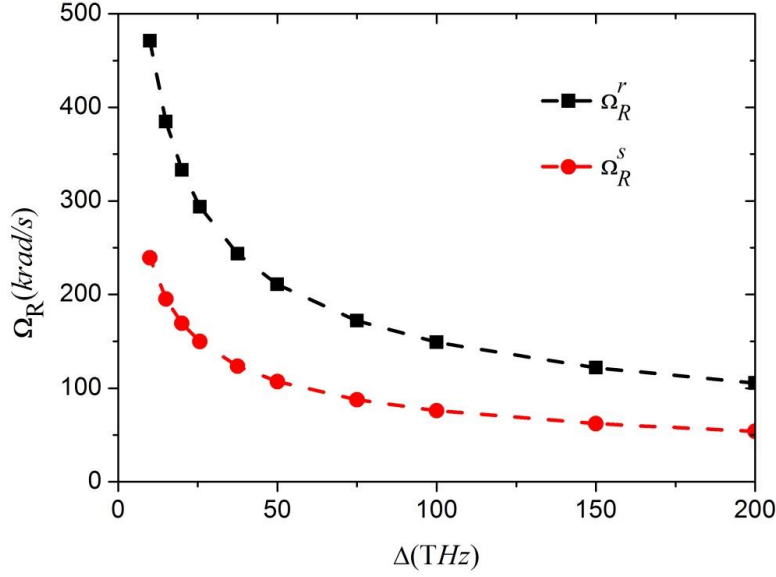

FIG. 12. The variation of $\Omega_R^r$ and $\Omega_R^s$ as a function of the detuning $\Delta$ for a $^{85}$Rb atom initially at rest, where $l = 1$, $w_0 = 5\mu m$, and $P = 80 mW$.

## VI. CONCLUSIONS

In our work we investigated the motion of a two-level atom inside a rotating HOT. The rotation is induced when there is a frequency difference between the two beams that are superposed to generate the HOT and can be clockwise or counter-clockwise depending on the sign of the frequency difference. An ultracold atom with initial velocity almost zero can be guided along the rotating helical tube thanks to an induced motion along the radial direction due to an initial centrifugal force and an induced motion along the azimuthal direction due to the coupling with the radial motion, which gives rise to an initial Coriolis force.

In general the atomic motion inside the rotating HOT is of an oscillatory type. It is characterized by a global oscillation along the HOT axis and a local oscillation in the radial direction. The turning points of the global oscillation can be manipulated by the magnitude and direction of the HOT angular frequency $\Omega_R$. To be able to elevate the atom along the HOT axis we must release the atom from the global oscillation while simultaneously it remains trapped in the radial direction so it cannot escape from the optical tube. As our numerical work showed, this can be achieved when the HOT angular frequency has a certain range of values $|\Omega_R^s| < |\Omega_R| < |\Omega_R^r|$. The values of $\Omega_R^s$ and $\Omega_R^r$ can be controlled and manipulated by changing the dipole potential parameters such as the orbital angular momentum number $l$, the beam waist $w_0$, the beam power and the detuning from the atomic resonance.

Furthermore, the ultracold atom can be guided either upwards or downwards depending on the direction of the HOT rotation. It is found that if the HOT rotates counter-clockwise $\Omega_R > 0$ (clockwise $\Omega_R < 0$), then clockwise angular momentum (counter-clockwise) will be transferred to the atom which will let the atom move along the left–handed helical tube upwards (downwards). This result is consistent with the work of Bekshaev et al [18] where the net orbital angular momentum that is transferred by the HOT is opposite to the direction of the HOT rotation.

In the static left-handed HOT case we found that a trapped atom with non-zero initial velocity will elevate upwards to a specific turning point (with fixed values of the dipole potential's parameters) due to the coupling between the motions in the axial and azimuthal directions [13]. However, if the same atom with the same initial conditions is trapped in a rotating left-handed HOT (with the same values for the dipole potential parameters), then the position of the turning point can be manipulated by the value and direction of the angular velocity $\Omega_R$ of the HOT. Additionally, the effective frequency of the trapped atom oscillations can be manipulated by changing only the value of the angular velocity $\Omega_R$ of the HOT.

When the HOT rotates with $\Omega_R^s$, the value of the effective global frequency $\omega_s'$ drops to zero. In this case, there are not anymore oscillations of the atom along the $z$-direction. The rotating HOT provides the trapped atom with enough energy to escape from oscillations and start elevating. Furthermore, if the HOT rotates faster, up to $\Omega_R^r$, the value of the effective radial frequency $\omega_r'$ drops to zero. In this case, there are no more oscillations of the atom along the $r$-direction and this means that the atom escapes from the optical potential tubes.

As a conclusion if the HOT is rotated within a range $|\Omega_R^s| < |\Omega_R| < |\Omega_R^r|$ then the rotating HOT acts as an Optical Archimedes' Screw (OAS). The values of $\Omega_R^s$ and $\Omega_R^r$

and the range between these values can be controlled and manipulated by changing the dipole potential parameters such as the beam power, the detuning, the orbital angular momentum number $l$ and the beam waist $w_0$.